\documentclass{ws-procs975x65}

\def\Z#1{_{\lower2pt\hbox{$\scriptstyle#1$}}}

\begin{document}

\title{Black Holes, Entropy Bound and Causality Violation}

\author{I.~P.~NEUPANE}

\address{Department of Physics and Astronomy, University of Canterbury,\\
Private Bag 4800, Christchurch 8041, New Zealand\\
$^*$E-mail: ishwaree.neupane@canterbury.ac.nz}

\begin{abstract}

The gravity/gauge theory duality has provided us a way of studying
QCD at short distances from straightforward calculations in
classical general relativity. Among numerous results obtained so
far, one of the most striking is the universality of the ratio of
the shear viscosity to the entropy density. For all gauge theories
with Einstein gravity dual, this ratio is $\eta/s=1/4\pi$.
However, in general higher-curvature gravity theories, including
two concrete models under discussion - the Gauss-Bonnet gravity
and the $({\rm Riemann})^2$ gravity - the ratio $\eta/s$ can be
smaller than $1/4\pi$ (thus violating the conjecture bound), equal
to $1/4\pi$ or even larger than $1/4\pi$. As we probe spacetime at
shorter distances, there arises an internal inconsistency in the
theory, such as a violation of microcausality, which is correlated
with a classical limit on black hole entropy.

\end{abstract}

\keywords{AdS/CFT, higher-derivative black holes, entropy bound,
causality violation}

\bodymatter

\section{Introduction}

According to the famous dictum of gravity/gauge theory duality or
Maldacena's Anti de Sitter (AdS) conformal field theory (CFT)
correspondence~\cite{Malda99a}, an AdS space with a black hole is
dual to a field theory at a finite temperature. AdS/CFT has been
an excellent tool to study not only strongly coupled gauge
theories at a large 't Hooft coupling limit but also to study
hydrodynamic properties of a certain class of boundary CFTs at a
finite temperature. One of the highly celebrated results over the
last decade is that, for a large class of four-dimensional CFTs,
the ratio of the shear viscosity $\eta$ to the entropy density $s$
is (in units $\hbar = k_B = 1$) given by~\cite{KSS}
\begin{equation}
\frac{\eta}{s}=\frac{1}{4\pi}.
\end{equation}
This result has been known to hold for all gauge theories with
Einstein gravity dual. This is due to the fact that in pure
Einstein gravity all black holes satisfy the Bekenstein-Hawking
entropy law or the area formula of black hole entropy
\begin{equation}
S=\frac{k_B c^3}{\hbar} \frac{A}{4 G_N},
\end{equation}
where $\hbar$ is the Planck constant, $k_B$ is the Boltzmann
constant, $A$ is the area of the (black hole) horizon
corresponding to the surface at $r=r_{+}$ and $G_N$ is the
Newton's constant $G_N$. In general gravity theories with higher
derivative or higher-curvature corrections, however, the ratio
$\eta/s$ can be different from $1/4\pi$. In the same context, a
violation of causality might occur in the boundary CFT when
$\eta/s$ is too low ($\ll 1/4\pi$). Here we shall argue that any
such a violation of micro-causality can be related to a violation
of black hole entropy bounds or a violation of certain laws of
black hole thermodynamics.

To show that this might be the case, and also to establish better
contact with QCD via AdS holography, in a holographic context
(see~\cite{Ish02} and references therein), one might like to
consider the following gravitational action
\begin{equation}\label{main-action}
I_{\rm g}=\frac{1}{16\pi G_N}\int d^{d} x\sqrt{-g}
\left[R-2\Lambda +{\alpha^\prime} L^2 \left(a R^2+ b
R_{\mu\nu}R^{\mu\nu}+c
R_{\mu\nu\lambda\rho}R^{\mu\nu\lambda\rho}\right) \right],
\end{equation}
where $\alpha^\prime$ is a dimensionless coupling and
$\Lambda\equiv -(d-1)(d-2)/2L^2$ is a bulk cosmological term. In
fact, the Regge slope or the coefficient $\alpha^\prime$ appearing
in the effective action~(\ref{main-action}) can be a complicated
function of some microscopic parameters of the quantum gravity
theory, which may even depend on the details of compactification
and on dilaton couplings. Nevertheless, in the simplest scenario
to be considered in this paper, one would assume that
$\alpha^\prime =\ell_P^2/L^2 \lesssim {\cal O} (1/10)$. Here one
might also note that, especially, in an AdS$_5$ space, the
coefficients of $R^2$ terms in a dual supergravity action are
determined by the central charges of a CFT in four dimensions.

In most of our discussions we will focus to the gravity sector in
a five-dimensional AdS space, for which we have from the AdS/CFT
correspondence, $G_N\equiv (\pi L^3/4 N_c^2)$ and $L = (4\pi g_s
N_c)^{1/4} \ell_s$, where $\ell_s$ is the string scale and $g_s$
the string coupling and Nc is the number of color charges or rank
of the gauge group. In the dual supergravity description, a small
$\alpha^\prime$ corresponds to the strong coupling limit, i.e.,
$\lambda\equiv g_{YM}^2 N_c\gg 1$, since $\alpha^\prime\sim
1/\sqrt{\lambda}$. The Gauss-Bonnet term is obtained by setting $a
= c = 1$ and $b =-4$ in (\ref{main-action}), for which there would
be no need to treat $\alpha^\prime$ as small, at least, for the
purpose of obtaining exact (black hole)
solutions~\cite{Wiltshire:85a,Ish04C}.

Recently, Kats and Petrov~\cite{Kats-etal} and Brigante et
al.~\cite{Brigante-etal} have shown that, for a class of CFTs in
flat space with Gauss-Bonnet gravity dual, the ratio $\eta/s$,
which reads
\begin{equation}
\frac{\eta}{s}=\frac{1}{4\pi}\left(1-\frac{2(d-1)\lambda_{GB}}{(d-3)}\right),
\end{equation}
where $\lambda_{\rm GB} \equiv (d-3)(d-4)\alpha^\prime$, can be
smaller than $1/4\pi$ for $\lambda_{\rm GB}>0$ and $d\ge 5$. Based
on the bulk causal structure of an AdS$_5$ black brane solution, one
may actually find an even more stronger bound for $\lambda_{GB}$,
that is,
\begin{equation}
\lambda_{\rm GB} < \frac{9}{100} \quad {\rm or},~ {\rm
equivalently}, ~~ \frac{\eta}{s}\ge \frac{16}{25}
\left(\frac{1}{4\pi}\right),
\end{equation}
which otherwise violates a microcausality in the dual CFT defined on
a flat space. Here we show that the critical value of $\lambda_{GB}$
beyond which the theory becomes inconsistent is related to the
entropy bound for a large class of AdS black holes. In fact, in the
holographic context, AdS black hole solutions with spherical and
hyperbolic event horizons allow much wider possibilities for
$\eta/s$~\cite{Ish:08a}.

It is quite plausible that in the presence of curvature terms like
$R^n$ with $n\ge 3$, which generate six and higher derivatives in
the metric, or the $({\rm Weyl})^4$ terms arising as
${\alpha^\prime}^3 R^4$ corrections to low energy effective action
of type IIB string theory, the ratio $\eta/s$ takes a value
slightly larger than $1/4\pi$ (see~\cite{Myers:09a} for a review).
However, in any consistent gravity models, interaction terms like
$\lambda L^4 R R_{abcd}R^{abcd}$, where $\lambda\sim \ell_P^4/L^4
\ll \alpha^\prime$, must be suppressed in a sensible derivative
expansion, implying that such corrections could arise only as
subleading or next-to-subleading terms. Here we show that when an
underlying theory admits a large violation of the conjectured KSS
bound $\eta/s\ge 1/4\pi$ or the ratio $\eta/s$ becomes too small,
then certain laws of thermodynamics can be violated in the same
limit. This effect can be seen also in terms of a violation of
causality in the boundary field theory. The bound $\eta/s\ge
1/4\pi$ may be restored only at weak couplings or in the limit of
large $N_c$, since in such cases the shear viscosity can grow
faster as compared to the entropy density.

\section{Gauss-Bonnet Gravity and Causality Violation}

For an AdS GB black hole, the entropy and Hawking temperature are
given by~\cite{Ish04C}
\begin{equation}\label{GB-entropy}
{\cal {\cal S}} = \frac{A}{4G_N}\left(1+ \frac{2(d-2)k
\lambda_{\rm GB} }{(d-4)x^2} \right),\quad
 T=\frac{(d-1)x^4+ k(d-3)x^2+ (d-5)k^2 \lambda_{\rm GB}}
 {4\pi L x(x^2+2 k \lambda_{\rm
GB})}
\end{equation}
(in units $c=\hbar=k_B=1$) where $x\equiv {r_{+}}/{L}$, $A \equiv
V_{d-2} r_{+}^{d-2}$, with $V_{d-2}$ being the unit volume of the
base manifold or the hypersurface ${\cal M}$. The entropy of a GB
black hole depends on the curvature constant $k$, whose value
determines the geometry of the event horizon ${\cal M} ={\rm
S}^{d-2}, {\rm I\!\, \! R}^{d-2}$ and ${\rm H}^{d-2}$,
respectively, for $k=+1, k=0$ and $k=-1$~\footnote{The GB term is
topological in $d=4$, especially, with a constant coupling,
$\alpha^\prime L^2={\rm const}$, so we take $d\ge 5$.}.
Especially, at the $k=-1$ extremal state with zero temperature,
\begin{equation}
{\cal S}|_{T\to 0} =\frac{V_3}{G_N} \frac{L^3}{2^{7/2}}
(1-12\lambda_{GB}), \quad E|_{T\to 0}=0.
\end{equation}
Thus, beyond a critical coupling $\lambda_{GB} > \lambda_{\rm
crit}$, the entropy ${\cal S}$ becomes negative, which indicates a
violation of cosmic censorship or the second law of the
thermodynamics. In the AdS$_5$ case, $\lambda_{\rm crit}=1/12$. This
critical value of $\lambda_{GB}$ above which the theory is
inconsistent nearly coincides with the bound $\lambda_{GB}<9/100$
required for a consistent formulation of a class of CFTs in a flat
space with Gauss-Bonnet gravity dual.

To be more specific, let us consider small metric fluctuations
$\phi=h_2^1$ around an AdS GB black hole metric~\footnote{There
can be three different modes of metric fluctuations, each of which
may be decoupled from others: the scalar, vector and tensor
fluctuations. For simplicity, and especially for the purpose of
calculating shear viscosity, one may study only the tensor
fluctuations $h_{x y}$, using $\phi$ to denote this perturbation
with one index raised $\phi = h_x^y$ and writing $\phi$ in the
basis $\phi(t, x_3, z) = \phi(z) e^{-iw t +i q x_3}$.}

\begin{equation}
ds^2=-\frac{r_+^2}{L^2} f(z) N_{\ast}^2 dt^2 +
\frac{L^2}{f(z)}dz^2  + \frac{r_+^2 z^2}{L^2} \left(\frac{d
x_{3}^2}{1-k x_{3}^2}+ x_{3}^2 \sum_{i=1}^{2} dx_i^2 + 2\phi(t,
x_{3} ,r) dx_1 dx_2\right),
\end{equation}
where $k=0, \pm 1$ and $N_{\ast}\equiv a
=\left[(1+\sqrt{1-4\lambda_{GB}}\,)/2\right]^{1/2}$ and $L$ is the
curvature of AdS$_5$ space. The scalar metric fluctuation (along
the $x_3$ direction)
\begin{equation}
\phi(t, x_{3}, z)=\int \frac{dw dq}{(2\pi)^3} \, \phi(z ;
\hat{k})\, e^{-iwt+iq x_3}, \quad \phi(z; -\hat{k})=\phi^*(z,
\hat{k}),
\end{equation}
(where $\hat{k}=(w, 0, 0,q)$) satisfies the following (linearized)
equation of motion
\begin{equation}\label{linear-eq}
K\,\partial_z^2\phi +\partial_zK\,\partial_z\phi + K_2 \phi =0.
\end{equation}
This structure is not affected by Maxwell type charges but it may
well be affected by dilatonic scalar charges and rotation
parameters. In the simplest scenario one may neglect both the
scalar charge and rotation parameters. In pure GB gravity defined
on an AdS$_5$ spacetime, we find $K=z^2 \tilde{f}
\left(z-{\lambda_{\rm GB}\partial_z {f}}\right)$, $K_2=\left(z^2
\tilde{w}^2/ N_*^2 f\right) \left(z-{\lambda_{\rm GB}\partial_z
f}\right) -z \left(1-\lambda_{\rm GB}
\partial_z^2
{f}\right)\left(\tilde{q}^2+2\tilde{k}\right)$ and
%%%
\begin{equation}\label{sol-fr}
f(z)=\tilde{k} +\frac{z^2}{2\lambda_{GB}} \left[1\pm
\sqrt{1-4\lambda_{\rm GB}+\frac{4\lambda_{\rm
GB}}{z^4}\left(1+\frac{k}{x^2}+\frac{\lambda_{\rm
GB}k^2}{x^4}\right)}\,\right],
\end{equation}
where $z\equiv r/r_+$, $x\equiv r_+/L$, $\tilde{k}=k/x^2$,
$\tilde{w}\equiv w L/x$ and $\tilde{q}\equiv qL/x$.
Eq.~(\ref{linear-eq}) may be solved with the following incoming
boundary conditions at the horizon
\begin{equation}
\phi(z; \hat{k})= a_{\rm in}(\hat{k}) \phi_{\rm in}(z; \hat{k})+
a_{\rm out}(\hat{k}) \phi_{\rm out}(z; \hat{k}), \quad a_{\rm
out}\equiv 0, \quad a_{\rm in}\equiv J(\hat{k}),
\end{equation}
where $J(\hat{k})$ is an infinitesimal boundary source for the
fluctuating field $\phi$. To the leading order in $\tilde{w}$, and
in the limit $\tilde{q}\to 0$, the solution is given
by~\cite{Ish04C}
\begin{equation}
\phi(z; \hat{k}) =J(\hat{k}) \left[1+\frac{i\tilde{w}}{4N_*} a^2
\sqrt{1-4\lambda_{\rm GB}}\left(\frac{1}{z^4}-\frac{4\lambda_{\rm
GB} \tilde{k}}{3(1+\sqrt{1-4\lambda_{\rm GB}})}\frac{1}{z^6}+
{\cal O}(z^{-8})\right)\right].
\end{equation}
By identifying $\phi\sim T_1^2$, where $T_1^2$ is a component of
the energy-momentum tensor on the field theory side, we may obtain
the retarded two-point Green's function. Indeed, GB black holes
admit a stable five-dimensional anti-de Sitter vacuum only if
$\lambda_{GB} < 1/4$~\cite{Ish02}. This result is independent of
the types of interactions the Gauss-Bonnet term can have with
matter fields (including gauge and scalar fields).

From the above result, we can easily see that the curvature on a
boundary may not affect the shear viscosity
\begin{equation}\label{main-result}
{\eta}=\frac{1}{16\pi G_N} \left(\frac{r_{+}^3}{L^3}\right)
(1-4\lambda_{\rm GB})
\end{equation}
obtained using the Kubo formula
\begin{equation}\label{Kubo-formula}
\eta = \lim_{w\to 0}
\frac{1}{2iw}\left[G^A_{12,12}(w,0)-G^R_{12,12}(w,0)\right] \equiv
\lim_{w\to 0} \frac{1}{w} {\rm Im} G^R_{12, 12} (w,0).
\end{equation}
As is evident, this formula relates $\eta$ to zero spatial
momentum ($q=0$), low frequency limit of the retarded two-point
Green's function
\begin{equation}
G^{R}_{12,12}(w,0)= -i \int d^4 x e^{iwt}\theta(t) \langle
[T_{12}(t,{\vec x}), T_{12}(0,{\vec 0})]\rangle,
\end{equation}
which satisfies $G^A(w, \vec{q})\equiv G^R(w,\vec{q})^*$. The
ratio $\eta/s$ is now modified as
\begin{equation}
\frac{\eta}{s} =\frac{1}{4\pi} \frac{(1-4\lambda_{\rm
GB})}{(1+6\tilde{k} \lambda_{\rm GB})}.
\end{equation}
The standard result in Einstein gravity, i.e. $\eta/s=1/4\pi$, is
obtained only at a fixed $\tilde{k}$, i.e. when $k=-1$ and
$x=r_+/L=\sqrt{3/2}$. The minimum of entropy density actually
occurs at $x=\sqrt{1/2}$, implying that
\begin{equation}\label{GB-extremal-s}
s=\frac{1}{G_N} \frac{1}{2^{7/2}}\left(1-12\lambda_{\rm
GB}\right), \quad \eta=\frac{1}{4\pi G_N}
\frac{1}{2^{7/2}}\left(1-4\lambda_{GB}\right).
\end{equation}
At this extremal state the shear viscosity is given by
\begin{equation}
\eta=\frac{1}{4\pi G_N} \frac{1}{2^{7/2}}\left(1-4\lambda_{\rm
GB}\right).
\end{equation}
The positivity of extremal entropy density implies that
$\lambda_{GB}\le 1/12$ and hence $\eta/s\ge 1/6\pi$ for the $k=0$
and $\eta/s< 5/12\pi$ for the $k=-1$ solutions. It is quite
remarkable that the lower bound found above, i.e. $\eta/s\approx
0.66/4\pi$, is similar to a lower value of $\eta/s$ found at some
relativistic heavy ion collision experiments~\cite{Molnar05}. The
bound $\eta/s> 0.09$ was found in~\cite{Brigante-etal} by
considering a Ricci flat horizon or assuming that the boundary
theory is defined on a flat space. Here we note that, in the
$k=+1$ case, the lower value of $\eta/s$ can be slightly stronger
than that in flat space. Especially, in a flat space, a
non-violation of causality requires the square of local speed of
graviton on a constant $z$-hypersurface to be less than unity,
i.e.
\begin{equation}
c_g^2 = 1- \left(\frac{5}{2}
-\frac{2}{1-4\lambda_{GB}}+\frac{1}{2\sqrt{1-4\lambda_{GB}}}\right)\frac{1}{z^4}
+{\cal O}(z^{-8}) <1
\end{equation}
or $\lambda_{GB} < 0.09$. This limit can be slightly altered by
the Maxwell term $F_{\mu\nu}F^{\mu\nu}$ and also by $F^4$ type
corrections to Maxwell fields~\cite{Ge-Sin,Cai08k}~\footnote{When
matter fields (including gauge field and scalar fields) are
coupled to higher-curvature terms, and as long as Einstein's field
equations contains at most second derivatives of the metric
component $g_{xy}$ as a function of $t, z, x_i$, which is the case
only with the Gauss-Bonnet gravity, then the perturbation of the
transverse gravitons can get decoupled from the fluctuations of
matter fields.}. Specifically, in the presence of a Maxwell charge
$q$, the ratio $\eta/s$ is given by $\eta/s=\frac{1}{4\pi}\left(1-
4\lambda_{\rm GB} + 2 Q \lambda_{\rm GB}\right)$, where $Q \equiv
\frac{q^2 L^2}{r_+^6}$. In the standard case of an non-extremal
black hole one may be required to satisfy two separate conditions
$Q< 2$ and $\lambda_{\rm GB}< 1/24$, leading to the bound $\eta/s>
\frac{1}{4\pi} \times \frac{5}{6}$ as first reported
in~\cite{Ge-Sin}.

For the consistency of a dual Gauss-Bonnet gravity action, and
also from a viewpoint of classical stability of GB black holes,
one is required to take
\begin{equation}
\lambda_{GB}^{d=6}\lesssim 0.1380 \quad {\rm and} \quad
\lambda_{GB}^{d=7}\lesssim 0.1905.
\end{equation}
respectively, in the AdS$_6$ and the AdS$_7$ cases.

Next let us briefly discuss the results in the $({\rm Riemann})^2$
gravity, which is obtained by setting $a=0=b$ and $d=5$ in
Eq.~(\ref{main-action}). In this theory, we
find~\cite{Ish:08a,Ish04C}~\footnote{We believe the result
in~\cite{Brigante-etal}, i.e.
$\eta/s=\frac{1}{4\pi}\left(1-4\lambda_{Riem}+\cdots\right)$
should be corrected by a factor of 2: the source of this
difference is that in $({\rm Riemann})^2$ gravity the entropy
function is increased by a factor of $(1+4\lambda_{\rm Riem})$, in
the case of a Ricci-flat horizon, which was however not considered
there.}
\begin{equation}\label{main-SqRm}
\frac{\eta}{s}=\frac{1}{4\pi} \left(\frac{1-4\lambda_{\rm
Riem}}{1+4\lambda_{\rm Riem}}\right)\approx \frac{1}{4\pi}
\left(1-8 \lambda_{\rm Riem}+\cdots \right)=\frac{1}{4\pi}
\left(1-\frac{1}{N_c}+\cdots \right).
\end{equation}
This result shows that the KSS bound $\eta/s \ge 1/(4\pi)$ can be
violated for a finite $N$, unless that the contribution to
$\eta/s$ coming from next-to-leading order terms, such as
$(\rm{Weyl})^4$ terms, are of a comparable magnitude to that of
the $R^2$ terms. This scenario is, however, almost unlikely since
the couplings associated with the cubic and higher powers in
Riemann tensors, such as $\lambda L^4 R R_{abcd}R^{abcd}$ with
$\lambda\sim \ell_P^4/L^4 \ll \alpha^\prime$, must be suppressed
in a sensible derivative expansion.

Notice that the limit $\lambda_{\rm Riem}< 1/8$, as implied by
Eq.~(\ref{main-SqRm}) and required for positivity of $\eta/s$, is
the same as implied by the positivity of extremal entropy of a
$({\rm Riemann})^2$-corrected AdS black hole. This result is
consistent with the following observation. The conformal anomaly
of a four-dimensional CFT can be identified by considering the
theory in a curved spacetime and writing
\begin{equation}
16\pi^2 {\langle T_\mu\,^\mu\rangle}_{CFT}= - c_1\,E_4 - c_2\,
W_4,
\end{equation}
where $E_4\equiv
R_{\mu\nu\lambda\sigma}R^{\mu\nu\lambda\sigma}-4R_{\mu\nu}R^{\mu\nu}+R^2$
and $W_4\equiv
R_{\mu\nu\lambda\sigma}R^{\mu\nu\lambda\sigma}-2R_{\mu\nu}R^{\mu\nu}+R^2/3$
are, respectively, the four-dimensional Euler density and the
square of the Weyl curvature, and $c_1$, $c_2$ are the two central
charges of a dual CFT. As explained by Hofman and Maldacena
in~\cite{Hofman:08a} (see also~\cite{Myers:09a}), for the
consistency of a low energy action with sensible derivative
expansion, one may be required to consider CFTs for which
\begin{equation}
| 1-\frac{c_1}{c_2}| \lesssim {\cal O}(1/10).
\end{equation}
In fact, the gravity dual of a class of CFTs for which
$\lambda_{Riem}\simeq \frac{1}{8}
\left(1-\frac{c_1}{c_2}\right)\ne 0$, must be defined in a curved
background, so the choice $k=0$ may not be very physical, at
least, in the presence of $({\rm Riemann})^2$ type corrections.

Taking into account all three possibilities for the boundary
topology that $k=0$ or $k=\pm 1$, and demanding that
$\lambda_{Riem}\simeq \frac{1}{8c_2}(c_2-c_1)<1$, we find
\begin{equation}
0 < \frac{\eta}{s} \le \frac{3}{2} \left(\frac{1}{4\pi}\right).
\end{equation}
It is not known yet whether either of these limits applies to
nuclear matters at extreme densities and temperatures, or heavy
ion collision experiments, but it would be interesting to know
anything specific to the universality of the result $\eta/s\approx
1/4\pi$ through numerical hydrodynamic simulations of data from
high energy experiments, including the relativistic heavy ion
collision (RHIC) and large hadron collider.

\section{Conclusion}

Recent developments in gravity/gauge theory duality,
supersymmetric field theories and black hole mechanics have shown
that AdS black holes are excellent objects to study the properties
of strongly coupled gauge theories. In this note we have studied
the limits on black hole entropy and shear viscosity for the
simplest class of higher curvature-corrected black hole solutions
defined on AdS spaces. We also considered the causal problem on
the boundary field theory to see what kind of constraints we can
get on the ratio $\eta/s$ (where $\eta$ is the shear viscosity and
$s$ is the entropy density) so as to keep the theory
phenomenologically viable and internally consistent.

It has been known for quite sometimes that the limit $\lambda_{GB}
< 1/4$ on the Gauss-Bonnet coupling can be viewed as a classical
limit of a consistent theory of quantum
gravity~\cite{Wiltshire:85a,Ish04C}. Recent results coming from
the studies of viscosity bound violation in higher curvature
gravity~\cite{Brigante-etal,Ish:08a} have shown that the bound on
the shear viscosity of any fluid in terms of its entropy density
may be saturated, i.e. $\eta/s=1/4\pi$, by all gauge theories but
only at large 't Hooft coupling, as this limit correspond to the
cases where all higher-order curvature contributions are absent.
Nevertheless, this bound is naturally in immediate threat of being
violated in the presence of generic higher derivative and
higher-order curvature corrections to the Einstein-Hilbert action.

It is important to note that by tuning of the Gauss-Bonnet
coupling or the generic higher derivative couplings, the ratio
$\eta/s$ can be adjusted to a small positive value. Causality
violation might take place when $\eta/s$ is too low, i.e.
$\eta/s\ll 1/4\pi$.

We have shown that limits on curvature coupling can be imposed by
demanding the positivity of extremal black hole entropy or by
keeping boundary causality in tact, or both. We have not found any
obvious explicit bound on $\lambda_{GB}$ from the thermodynamics
of spherically symmetric AdS Gauss-Bonnet black holes. This could
however arise as a consequence of causality violation of a
boundary CFT. The critical value of $\lambda_{GB}$ beyond which
the theory becomes inconsistent is found to be related to the
entropy bound for an AdS GB black hole with a hyperbolic or
Euclidean anti-de Sitter event horizon. This remark applies also
to the $({\rm Riemann})^2$ gravity. Some other inconsistencies of
higher-derivative gravity, such as an appearance of tacyonic mode,
or semi-classical instability at short distances, can also be
related to classical limits on black hole entropy and viscosity
bounds.

We conclude with a couple of remarks. In the presence of a Maxwell
type charge $q$, the ratio $\eta/s$ is generally modified, which
is given by $4\pi \left(\eta/s\right) = 1-4
\lambda_{GB}\left(1-Q/2\right)$, where $Q\equiv q^2 L^6/r_+^2$. In
the extremal limit ($Q\to 2$), we put a restriction on the
coupling $\lambda_{GB}$ such that $\lambda_{GB}\le 1/24$, which
guarantees that the gravitational potential of a black brane is
positive and bounded. The KSS bound $\eta/s\ge 1/4\pi$ is
saturated only in the extremal limit, while in general it is
violated also by charged Gauss-Bonnet black brane solutions. The
bound $4\pi(\eta/s)\ge 5/6$ found in~\cite{Ge-Sin} with a nonzero
charge is stronger than for pure Einstein-Gauss-Bonnet gravity in
flat space, namely $4\pi (\eta/s)\ge 2/3$. Our analysis showed
that in the AdS$_5$ case, the entropy of a GB black hole could be
negative for $1/12< \lambda_{GB} <1/4$, leading to a possible
violation of unitarity in this range. As shown recently
in~\cite{Cai:2009zv}, a scalar dilaton field coupled to the
Gauss-Bonnet can have an additional non-trivial contribution to
the ratio $\eta/s$: a scalar charge generally lowers the KSS
bound. In some more recent
papers~\cite{Cai:2009zv,Banerjee:2009wg,Mia:2009wj,Matsuo:2009yu},
there have appeared new examples, including string theory
constructions, where the standard result of Einstein gravity,
namely $\eta/s=1/4\pi$, is modified in a modest way.

In conclusion, we have explored several interesting connections
between the bulk causality and limits on black hole entropy,
specific to the modified Gauss-Bonnet and the $({\rm Riemann})^2$
gravity models.

\vspace{1.0ex}
\begin{flushleft}
\large\bf Acknowledgments
\end{flushleft}

I would like to acknowledge a fruitful collaboration and helpful
discussions with Naresh Dadhich. I am grateful to the organizers
of the JGRG18 (Japan) and the conference on Particle Physics,
Astrophysics and Quantum Field Theory: 75 Years since Solvay
(Singapore) for their hospitality. This work was supported by the
New Zealand Foundation for Research, Science and Technology Grant
No. E5229 and also by Elizabeth Ellen Dalton Grant No. 5393.

\end{document}